\documentclass[twocolumn,aps,prx,superscriptaddress,longbibliography]{revtex4-2}
\usepackage{amsmath}
\usepackage{amssymb}
\usepackage{graphicx}
\usepackage{color}
\usepackage[dvipsnames]{xcolor}
\usepackage{tabularx}
\usepackage{float}
\usepackage{array}
\usepackage{multirow}
\usepackage{arydshln}
\usepackage{booktabs}
\usepackage{graphicx}
\usepackage{epstopdf}
\newcommand{\be}{\begin{equation}}
\newcommand{\ee}{\end{equation}}
\newcommand{\bea}{\begin{eqnarray}}
\newcommand{\eea}{\end{eqnarray}}
\newcommand{\bse}{\begin{subequations}}
\newcommand{\ese}{\end{subequations}}
\newcommand{\CaMnP}{CaMn$_2$P$_2$}
\newcommand{\SrMnP}{SrMn$_2$P$_2$}

\newcommand{\TN}{$T_{\rm N}$}

\newcommand{\bfq}{\mathbf{q}}
\newcommand{\bfr}{\mathbf{r}}
\newcommand{\bfrr}{\mathbf{R}}

\newcommand{\bfqq}{\mathbf{Q}}
\newcommand{\bfss}{\mathbf{S}}

\definecolor{go_green}{rgb}{0.13, 0.55, 0.13}

\usepackage{hyperref}
\hypersetup{
    colorlinks=true,
    linkcolor=Blue,
    urlcolor=Maroon,
    citecolor=Mulberry,
    pdftitle={Spiral Spin Liquid State in the Corrugated Honeycomb Lattice in CaMn$_2$P$_2$},
    pdfauthor={Farhan Islam, Nikita D. Andriushin, Tha\'is V.~Trevisan, Santanu Pakhira, Simon X.~M.~Riberolles, Zachary Morgan, Arianna Minelli, Feng Ye, David C. Johnston, Robert J. McQueeney, Peter P. Orth, David Vaknin}
}

\begin{document}

\author{Farhan Islam}
\affiliation{Ames National Laboratory, Ames, Iowa 50011, USA}
\affiliation{Department of Physics and Astronomy, Iowa State University, Ames, Iowa 50011, USA}
\affiliation{Neutron Scattering Division, Oak Ridge National Laboratory, Oak Ridge, Tennessee 37831, USA}

\author{Nikita D.~Andriushin}
\affiliation{Institut für Festkörper- und Materialphysik, Technische Universität Dresden, D-01069 Dresden, Germany} 

\author{Tha\'is V.~Trevisan}
\affiliation{S\~{a}o Carlos Institute of Physics, University of S\~{a}o Paulo, 13560-970, S\~{a}o Carlos, S\~{a}o Paulo, Brazil.}

\author{Santanu~Pakhira}
\affiliation{Ames National Laboratory, Ames, Iowa 50011, USA}
\affiliation{Department of Physics and Astronomy, Iowa State University, Ames, Iowa 50011, USA}
\affiliation{Department of Physics, Maulana Azad National Institute of Technology, Bhopal 462003, India}

\author{Simon X.~M.~Riberolles}
\affiliation{Ames National Laboratory, Ames, Iowa 50011, USA}

\author{Zachary Morgan}
\affiliation{Neutron Scattering Division, Oak Ridge National Laboratory, Oak Ridge, Tennessee 37831, USA}

\author{Arianna Minelli}
\affiliation{Neutron Scattering Division, Oak Ridge National Laboratory, Oak Ridge, Tennessee 37831, USA}

\author{David~C.~Johnston}
\affiliation{Ames National Laboratory, Ames, Iowa 50011, USA}
\affiliation{Department of Physics and Astronomy, Iowa State University, Ames, Iowa 50011, USA}

\author{Robert J. McQueeney}
\affiliation{Ames National Laboratory, Ames, Iowa 50011, USA}
\affiliation{Department of Physics and Astronomy, Iowa State University, Ames, Iowa 50011, USA}

\author{Feng Ye}
\affiliation{Neutron Scattering Division, Oak Ridge National Laboratory, Oak Ridge, Tennessee 37831, USA}

\author{Peter P. Orth}
\affiliation{Ames National Laboratory, Ames, Iowa 50011, USA}
\affiliation{Department of Physics and Astronomy, Iowa State University, Ames, Iowa 50011, USA}
\affiliation{Department of Physics, Saarland University, 66123 Saarbr\"ucken, Germany}

\author{David Vaknin}
\affiliation{Ames National Laboratory, Ames, Iowa 50011, USA}
\affiliation{Department of Physics and Astronomy, Iowa State University, Ames, Iowa 50011, USA}

\title{Spiral Spin Liquid State in the Corrugated Honeycomb Lattice of CaMn$_2$P$_2$}
\date{\today}

\begin{abstract}
\CaMnP\ exemplifies the realization of a frustrated $J_1$-$J_2$-$J_3$  Heisenberg model of a corrugated honeycomb magnetic lattice. Previous studies show that below the N\'eel temperature (\TN), the system forms a cycloidal $6\times 6$ $ab$-plane magnetic unit cell that conforms with various magnetic space groups. Here, we present single-crystal neutron-diffraction studies across expansive reciprocal-space volumes, confirming the cycloidal magnetic structure while uncovering further distinctive features. We find evidence for three magnetic domains, the analysis of which narrows the possible magnetic model structures. At \TN, the insulator exhibits a sharp phase transition, above which the spin structure transforms into a spiral spin liquid state, evident via a continuous ring of scattering with degenerate wavevectors corresponding to a collection of short-range spiral spin configurations. These degenerate states emerge as thermal fluctuations effectively reduce the $J_3$ interaction. The integration of experimental, theoretical, and real-space simulation results reveals the intricate balance of exchange interactions ($J_1$-$J_2$-$J_3$) that stabilizes the ground-state magnetic structure and drives the emergence of a sought-after $U$(1)-symmetric spiral spin-liquid state with easy-plane anisotropy above the transition temperature.

\end{abstract}

\maketitle

\section{Introduction}
Competing near-neighbor spin interactions in magnetic systems can result in frustration and lead to a large number of magnetic configurations being energetically degenerate~\cite{Lacroix2011}. The presence of accidental degeneracies largely enhances the effects of quantum and thermal fluctuations and triggers the emergence of exotic phenomena and states such as spin liquids, emergent gauge fields, and topological order~\cite{Balents2010, Kitaev2006, Castelnovo2012}. Canonical spin model examples with an extensive ground state degeneracy are geometrically frustrated Ising or Heisenberg models with corner-shared triangular motifs such as the two-dimensional kagome~\cite{Kano1953, Moessner2000, Chalker1992} or the three-dimensional pyrochlore lattice~\cite{Castelnovo2012, Moessner1998}. Subextensive degeneracies are also common, for example, in frustrated, continuous spin models in two-dimensions (2Ds). They appear close to phase boundaries between classical ground states and manifest in a line of zero modes in the spin-wave spectrum. Examples are Heisenberg models on the square and honeycomb lattice with magnetic interactions that extend beyond the nearest neighbors~\cite{Rastelli1979, Chandra1988, Fouet2001}. These degeneracies can suppress conventional magnetic order and lead to the emergence of quantum spin liquid phases for low spin-$S$ systems, for example, on the square lattice at $J_2 \approx J_1/2$~\cite{Chandra1988, Jiang2012} and on the honeycomb lattice at $J_2 \approx J_1/6$~\cite{Fouet2001, Albuquerque2011, Gong2013, Reuther2011a}. 

A distinct class of systems with subextensive ground state degeneracy are spiral spin liquids (SSLs)~\cite{Bergman2007}. These classical spin liquids are cooperative paramagnets that exhibit a continuous degeneracy of single-$Q$ spiral ground states, {\it i.e.}~each ground state exhibits spiral magnetic order that is characterized by a single wavevector $\bfqq$. The collection of spiral wavevectors $\bfqq$ forms a low-dimensional manifold in momentum-space, which takes the form of a ring in some systems such as for extended spin models on honeycomb and square lattices~\cite{Okumura2010, Mulder2010a, Seabra2016, Yao2021}. 
This continuous degeneracy of spiral ground states leads to intriguing fluctuation properties. Unlike in spin-ice systems, where different states in the degenerate manifold can be explored via local spin flips, low-energy spin fluctuations in SSLs are spatially nonlocal and involve a global rotation of all the spins. Spin fluctuations can be described by a coarse-grained, position-dependent ordering wavevector $\bfqq(\bfr)$ that explores the degenerate manifold. If the degenerate manifold is highly symmetric (shaped as a circle, for example), order-by-disorder effects are suppressed to very low temperatures~\cite{Okumura2010, Mulder2010a}. In addition, if the degenerate manifold takes the form of a circle, the system exhibits an emergent $U(1)$ symmetry in momentum-space. Then, it was recently shown that there exist topological vortex excitations in momentum-space ({\it i.e.}~windings of $\bfqq(\bfr)$ around the degenerate circle) whose dynamics can be described by an elasticity theory that is equivalent to a tensor gauge theory of fractons with a generalized Gauss law~\cite{Yan2022a, Gonzalez2024}. 
This opens up the exciting possibility of observing the Kosterlitz-Thouless universality of these exotic vortex excitations~\cite{Gonzalez2024}. Since these vortices are only topological excitations if the spins are confined to a two-dimensional plane in spin space, a sufficiently large easy-plane anisotropy is required to realize this phenomenon. 
Finally, the predicted phase diagrams of SSLs in the presence of an external magnetic field are extremely rich and includes double and triple-Q states~\cite{Okubo2012,Shimokawa2019a} as well as an exotic ``ripple state" that hosts a single momentum-space vortex~\cite{Shimokawa2019}. The ripple state only emerges in the presence of $U(1)$ symmetry and for open boundary conditions~\cite{Gonzalez2024}.

Although many theoretical predictions about SSLs exist, only a few material realizations of SSLs are known. The characteristic experimental feature of a SSL is the emergence of continuous contours showing a high scattering intensity in momentum-space, which can be directly observed using neutron-scattering. 
The SSL regime typically occurs above a characteristic temperature, signaling that this state is entropically driven. The SSL state was first predicted for a diamond magnetic lattice commonly encountered in the $A$ site sublattice of spinel structures~\cite{Bergman2007}, and was later observed in MnSc$_2$S$_4$~\cite{Gao2017} and in CoAl$_2$O$_4$~\cite{Zaharko2011}. Recent high-resolution and diffuse magnetic neutron-scattering of polycrystalline LiYbO$_2$ showed continuous rings that can be interpreted as an SSL state on an elongated diamond-lattice~\cite{Graham2023}, and a SSL has been predicted for NiRh$_2$O$_4$~\cite{Chamorro2018,Chen2017, Buessen2018}. It has also been observed in the spinel MgCr$_2$O$_4$~\cite{Bai2019} with local magnetic moments occupying the $B$ pyrochlore sublattice. The degenerate spiral-$Q$ manifold in spinels is two-dimensional and highly anisotropic. Further, a SSL has been observed in the kagome bilayer material Ca$_{10}$Cr$_7$O$_{28}$~\cite{Balz2016,Balz2017,Pohle2021} and suggested to occur in the triangular bilayer system Ba$_3$NiSb$_2$O$_9$~\cite{Cheng2011, Chen2012}. Finally, recent neutron-scattering experiments in the van der Waals honeycomb Heisenberg-like magnet FeCl$_3$ have revealed the presence of continuous rings of high scattering intensity surrounding Bragg reflections, exemplifying a SSL with an approximate $U(1)$ symmetry in momentum-space~\cite{Gao2022}. 

Even with these few material realizations known, there remains an urgent need to discover new materials, in particular with a high-degree of $U(1)$ symmetry in momentum-space and easy-plane ($XY$) anisotropy, which would allow to experimentally explore the exotic physics driven by momentum vortices and their fate in magnetic fields and other external perturbations.

Here, we report neutron-scattering results that reveal a $U(1)$ symmetric SSL state in CaMn$_2$P$_2$ with a high degree of circular symmetry, substantial easy-plane anisotropy and a large transition temperature $T_{\text{N}} = 70$~K. We observe the characteristic ring of scattering above $T_{\text{N}}$ to extend over a wide temperature range beyond $T=120$~K. The SSL is borne out of a complex magnetic cycloidal ground state that we have explored previously~\cite{Islam2023a}. The magnetic Mn$^{2+}$ moments in \CaMnP\ are arranged in a corrugated honeycomb lattice and we account for our experimental observations using a previously established theoretical description of the material in terms of a highly frustrated $J_1$-$J_2$-$J_3$-$D_z$-$J_c$ spin model that considers frustrated and extended in-plane interactions and ferromagnetic coupling between honeycomb layers. Our detailed theory-experiment comparison places the material close to the maximally frustrated parameter regime of antiferromagnetic $J_2/J_1 \approx 0.25 \gg J_3/J_1$, where the degenerate manifold of single-Q wavevectors takes the shape of a highly-symmetric ring. Together with the experimental observation of a strong easy-plane anisotropy~\cite{Sangeetha2021}, we thus identify \CaMnP\ as an ideal material to further experimentally study the exotic fluctuation-driven phenomena associated with the presence of momentum-space vortices and their description in terms of a fracton tensor gauge theory~\cite{Yan2022a, Pretko2018}. 

\begin{figure}[h]
\centering
\includegraphics[width=.8\linewidth]{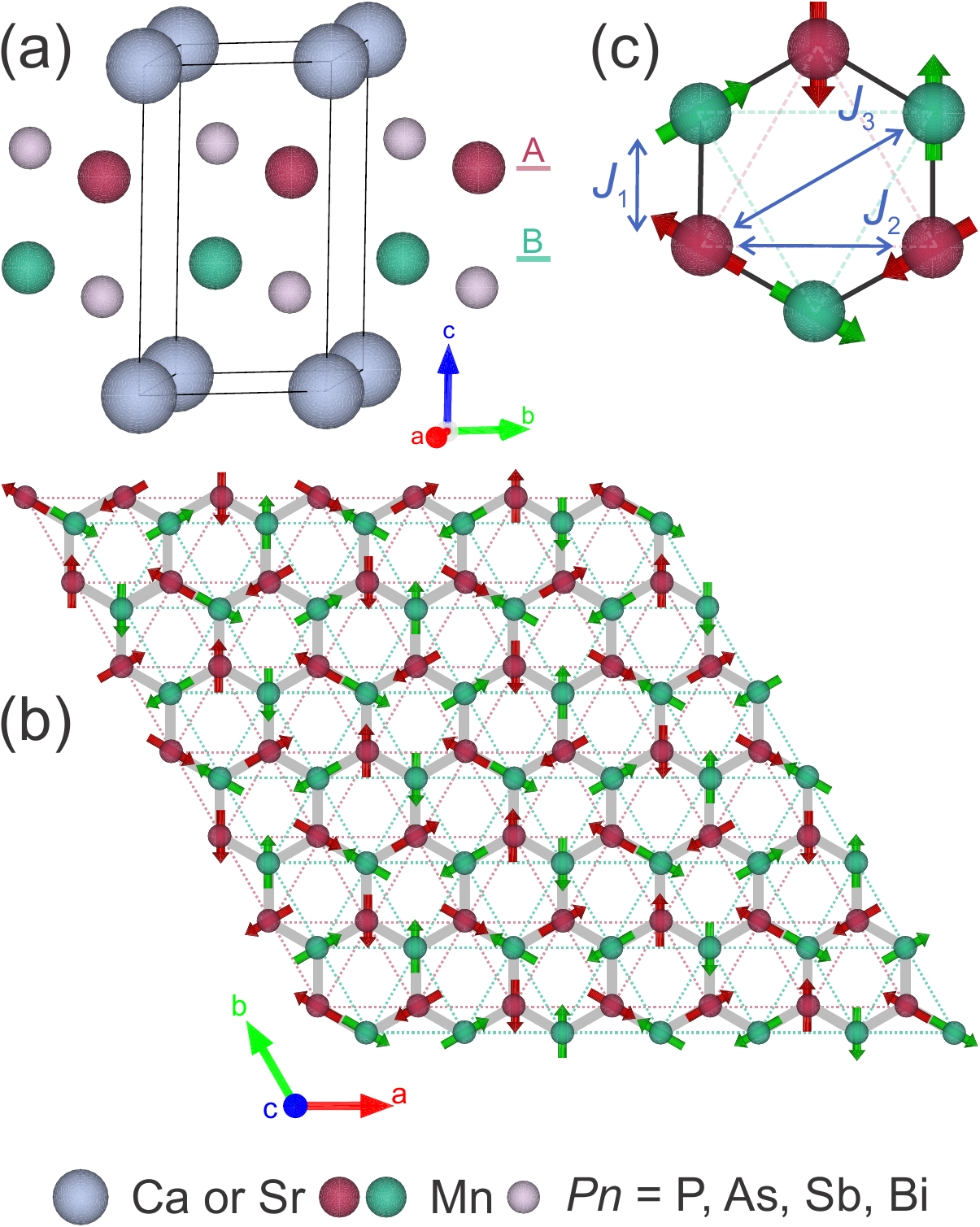}
\caption{(a) Chemical structure of $A$Mn$_2Pn_2$ ($A$ = Ca, Sr; $Pn$ = P, As, Sb, Bi) showing the trigonal Mn-bilayer ($A/B$ stacked) without intervening elements. (b)  Magnetic model structure with magnetic space group $P_Ac$ symmetry observed below \TN~\cite{Islam2023a}. It is constructed by creating a $6\times6$ $ab$-plane magnetic unit cell consisting of the corrugated honeycomb structure. The sites in green correspond to one trigonal layer (magnetic sublattice), and those in red to the other magnetic sublattice. (c) The panel indicates nearest-neighbor (NN) interactions ($J_1$), next-nearest-neighbor (NNN) interactions ($J_2$), and third-neighbor interactions ($J_3$).}
\label{Fig:Structure}
\end{figure}

\section{Experimental Details and Methods}
Single crystals of CaMn$_2$P$_2$ were grown in Sn flux as described previously \cite{Sangeetha2021}, and the crystals used in this study are from a similar growth batch. \CaMnP\ belongs to a group of Mn-based 122-type pnictides $A$Mn$_2Pn_2$ ($A$ = Ca, Sr; $Pn$ = P, As, Sb, Bi) \cite{Sangeetha2021,Sangeetha2016,Simonson2012a,Sangeetha2018,Das2017,Bridges2009,Gibson2015}. The chemical structure of $A$Mn$_2Pn_2$ depicted in Fig.~\ref{Fig:Structure}\hyperlink{Fig:Structure}{(a)} consists of Mn bilayers without intervening atoms. The projection of the two triangular Mn layers  onto the $ab$-planes shown in Fig.~\ref{Fig:Structure}\hyperlink{Fig:Structure}{(b)} forms a corrugated honeycomb lattice. Recent reports~\cite{Islam2023a} have shown that \CaMnP\ orders antiferromagnetically in a coplanar spiral magnetic state described by a \emph{commensurate} wavevector. It has also been reported that \SrMnP\ exhibits features of incommensurate magnetic order~\cite{Sangeetha2021,Brock1994}.

Neutron-diffraction measurements have been performed on the CORRELI diffractometer located at the Spallation Neutron Source at Oak Ridge National Laboratory. CORELLI is a quasi-Laue time-of-flight instrument implementing a correlation chopper to extract the elastic-scattering signal with an incident neutron wavelength-band between 0.7 and 2.9~\AA~\cite{Ye2018}. The two-dimensional detector array of CORELLI covers scattering angles from –22$^\circ$ to 148$^\circ$ in the horizontal plane and from –27° to 29° in the vertical direction. This allows collecting intensities over a large three-dimensional volume in reciprocal space that is achieved by rotating the single-crystal sample. The crystal was mounted on an Al pin at the tip of a closed-cycle refrigerator, providing temperature variation between 5 and 300 K. A $\sim 20$ mg crystal was mounted with the ($H$,$H$,0) and (0,0,$L$) in the scattering plane. We measured the lattice parameters to be $a=4.0942(3)$ and $c=6.8278(5) {\rm \AA}$ at base temperature ($T\approx5$ K) with the space group $P\bar{3}m1$ (no.~164). The experiments were conducted by first rotating the crystal 360$^\circ$ in 1.5$^\circ$ steps to survey the features in a large region of reciprocal-space; then, the data were collected for an extended time with particular sample orientations optimized for the selected reciprocal-lattice regions.
\section{Results and Discussion}
\label{sec:results_discussion}
\begin{figure*}
\centering
\includegraphics[width=.75\linewidth]{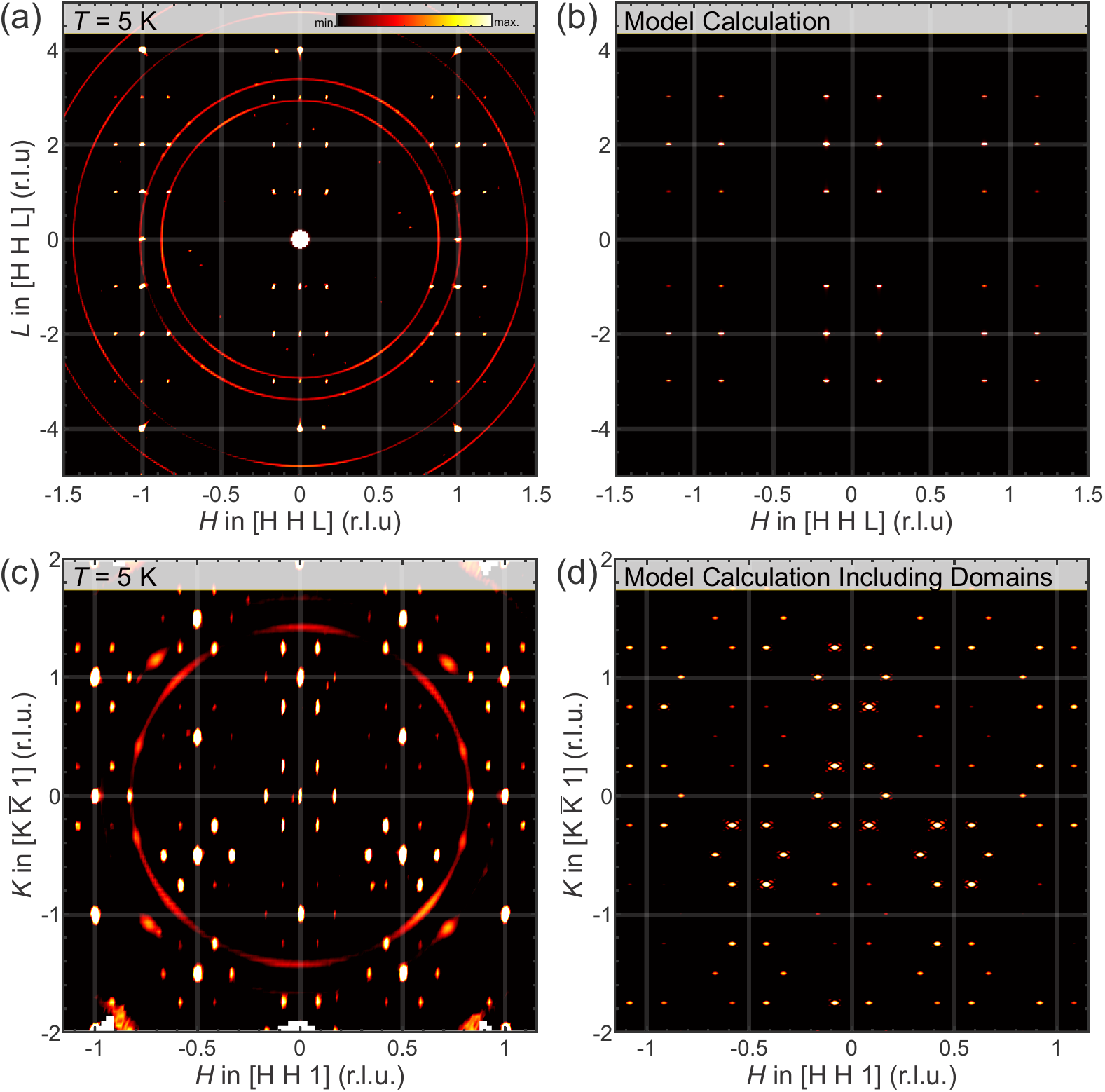}
\caption{(a) Experimental diffraction pattern in the ($H$,$H$,$L$) plane showing prominent magnetic Bragg reflections at ($n \pm 1/6$, $n \pm 1/6$, $L=\pm m)$~r.l.u. with integer $n,m \geq 0$ at $T=5$~K. The bright peaks at integer $H$ and $K$ are nuclear Bragg peaks present below and above $T_{\text{N}}$. 
(b) Results of a model calculation using the proposed magnetic structure with $P_Ac$ MSG (see Fig.~\ref{Fig:Structure}\protect\hyperlink{Fig:Structure}{(b)} showing good agreement with the experimental results. The simulation does not account for nuclear Bragg reflections. We note that the presence of three different domains does not lead to any qualitative difference in the $(H,H,L)$ scattering plane.  
(c) Experimental diffraction pattern in the ($H$,$K$,$1$) plane at $T=5$~K showing magnetic reflections at non-integer $H$ or $K$ values. (d) Model calculation results including three equally-populated magnetic domains can adequately capture the experimental data in the ($H$,$K$,$1$) plane.
}
\label{Fig:SF_base}
\end{figure*}
\subsection{Magnetic Structure and Domains}
\label{subsec:magnetic_structure}
\CaMnP\ exhibits a magnetically ordered state below $T_\text{N} = 70(1)$~K, which is described by an ordering vector $(H,K,L) = (\frac16,\frac16,0)$ r.l.u., leading to a $6 \times 6$ magnetic unit cell in the $ab$ plane~\cite{Islam2023a}. Previous neutron diffraction studies could not uniquely determine the specific magnetic space-group (MSG) and found that seven different ones with cycloidal magnetic structures equally conform to the observed diffraction patterns~\cite{Islam2023a}. Using the present diffraction data, we show that this can be narrowed down to three possible MSGs. One of them is the magnetic cycloidal structure with $P_Ac$ MSG that is illustrated in Fig.~\ref{Fig:Structure}\hyperlink{Fig:Structure}{(b)}. The magnetic arrangement is formed by rotating nearest-neighbor (NN) spins by 60$^\circ$ along the principal directions of a triangular sublattice (depicted by red arrows). The complementary sublattice is constructed by flipping all the spins in the first (red) sublattice by 180$^\circ$ (depicted by green arrows).  The $A/B$-stacked arrangement of these two sublattices [see Fig.~\ref{Fig:Structure}\hyperlink{Fig:Structure}{(c)}] results in the corrugated honeycomb lattice. 

The observed magnetic order can be rationalized using a frustrated $J_1$-$J_2$-$J_3$-$J_c$ Heisenberg model with easy plane anisotropy $D_z$ on ferromagnetically coupled honeycomb lattices
\begin{align}
\label{eq:Hspin}
H&=J_1\sum\limits_{\left\langle n,m\right\rangle_{1}}\mathbf{S}_n\cdot\mathbf{S}_m+J_2\sum\limits_{\left\langle n,m\right\rangle_{2}}\mathbf{S}_n\cdot\mathbf{S}_m  \\ & +J_3\sum\limits_{\left\langle n,m\right\rangle_{3}}\mathbf{S}_n\cdot\mathbf{S}_m + D_z \sum_n (S^z_n)^2 - J_c \sum\limits_{\left\langle n,m\right\rangle_{c}}\mathbf{S}_n\cdot\mathbf{S}_m \,.\nonumber
\end{align}
Here, $\langle n,m\rangle_i$ runs over $i$-th neighbors in a honeycomb layer and $\langle n,m \rangle_c$ denotes same sublattice sites in neighboring crystal unit cells. Besides, $\mathbf{S}_n = (S^x_n, S^y_n, S^z_n)$ denotes a classical spin of fixed length $|\mathbf{S}_n|=S$, well approximating the $L=0, S=5/2$ case of  Mn$^{2+}$ local moments. Honeycomb sites $n = (\mathbf{R}_n, \alpha)$ are labeled by a Bravais lattice vector $\mathbf{R}_n = n_1 \mathbf{a}_1 + n_2 \mathbf{a}_2 + n_3 \mathbf{a}_3$ with integers $n_i$ and a sublattice index $\alpha = A, B$. Here, the primitive vectors read $\mathbf{a}_1 = (a,0,0)$, $\mathbf{a}_2 = a(-\frac12, \frac{\sqrt{3}}{2},0)$ and $\mathbf{a}_3 = (0,0, a_c)$ with in-plane lattice constant $a=4.10$~\AA~and out-of-plane lattice constant $a_c=6.86$~\AA~\cite{Sangeetha2021} denoting the distance between different honeycomb layers.

For $J_2/J_1 > 1/6$ and small $J_3/J_1$, the classical ground state of Eq.~\eqref{eq:Hspin} is a single-Q spiral described as 
\begin{subequations}
    \begin{align}
        \bfss_{\rm A}(\bfrr_i) &= S \bigl( \sin(\mathbf{Q} \cdot \bfrr_i), \cos(\mathbf{Q} \cdot \bfrr_i), 0  \bigr) \label{eq:spin_parametrization_A} \\
        \bfss_{\rm B}(\bfrr_i) &= -S \bigl( \sin(\mathbf{Q} \cdot \bfrr_i + \phi), \cos(\mathbf{Q} \cdot \bfrr_i + \phi), 0\bigr) \label{eq:spin_parametrization_B}\,.
    \end{align}
\label{eq:spin_parametrizatio}
\end{subequations}
Here, we placed the spins into the $ab$ plane, consistent with experiment and sufficiently large easy-plane anisotropy $D_z > 0$. The angle $\phi+\pi$ describes the phase difference between the spins on $A$ and $B$ sites in the same unit cell. For $J_3 = 0$, there exists a continuous degeneracy of single-Q spiral states with wavevectors $\mathbf{Q} = (Q_a, Q_b) = \frac{Q_a}{2 \pi} \mathbf{G}_1 + \frac{Q_b}{2 \pi} \mathbf{G}_2$ that satisfy~\cite{Mulder2010a}
\begin{equation}
    \cos(Q_a) + \cos(Q_b) + \cos(Q_a + Q_b) = \frac12 \Bigl( \frac{J_1^2}{4 J_2^2} - 3 \Bigr) \,.
    \label{eq:wavevectors}
\end{equation}
Here, the in-plane reciprocal-lattice vectors of the honeycomb lattice read $\mathbf{G}_1 = (2 \pi, \frac{2 \pi}{\sqrt{3}a},0)$ and $\mathbf{G}_2 = (0, \frac{4 \pi}{\sqrt{3} a},0)$. The phase difference $\phi$ in Eq.~\ref{eq:spin_parametrizatio}\hyperlink{eq:spin_parametrizatio}{(b)} is determined by~\cite{Mulder2010a}
\begin{equation}
    \sin(\phi) = 2 J_2 \Bigl[ \sin(Q_b) + \sin(Q_a + Q_b) \Bigr] \,.
    \label{eq:phase_diff}
\end{equation}
As long as $1/6<J_2/J_1 < 0.5$, the manifold of degenerate wavevectors is a ring-like contour around the $(H,K,L)= (0,0,0)$ point. The ring contains the experimental ordering wavevector $H = K = \frac16$ for the highly frustrated spin exchange ratio $J_2/J_1 = 0.25, J_3 = 0$. Nonzero, small $J_3$ selects six, symmetry related ordering wavectors out of the degenerate manifold. The wavevector $H=K=\frac16$ is one of them for antiferromagnetic $J_3 > 0$ and $J_2 = \frac14(J_1 + 2 J_3)$~\cite{Islam2023a}. Since three of the magnetic wavevectors are related by $120^\circ$ rotations, we expect the formation of of three degenerate magnetic domains, each rotated by 120$^\circ$. Previous neutron-diffraction measurements have been limited to the $(H, H, L)$ plane and did not reveal these additional domains.

In the present study, neutron-diffraction experiments confirm the reported magnetic structure and the predicted domains. Fig.~\ref{Fig:SF_base}(a) displays the diffraction pattern in the ($0,0,L$) versus the ($H,H,0$) plane at base temperature ($T\approx5$~K). The pattern's bright spots correspond to nuclear ($H$ and $L$ integers) and magnetic ($H\pm1/6, H\pm1/6, L$) Bragg reflections.  The validation of previous findings is further supported by extending the scattering plane, providing a more comprehensive analysis. To rigorously assess the agreement between the observed and predicted magnetic diffraction patterns, we utilize the model structure depicted in Fig.~\ref{Fig:Structure}\hyperlink{Fig:Structure}{(b)} and follow the methodology for calculating intensities outlined in Ref.~\cite{Islam2023a}.
The calculated diffraction pattern in the $[H,H,L]$ plane is shown in Fig.~\ref{Fig:SF_base}\hyperlink{Fig:SF_base}{(b)}. It accurately reproduces the magnetic features around each nuclear Bragg reflection in the experimental data. Notably, our model calculations correctly reproduce the absence of magnetic Bragg reflections at $L=0$.
This robust agreement between the model and experimental data further establishes the validity of the proposed magnetic structure.

Figure~\ref{Fig:SF_base}\hyperlink{Fig:SF_base}{(c)} shows the experimental diffraction pattern in the  $(H,K,1)$ scattering plane, revealing six magnetic peaks surrounding integer $H$ and $K$ reciprocal-lattice points. Notably, no magnetic reflections are observed around the three lattice points $(1,0,1)$, $(0,-1,1)$, and $(-1,1,1)$. If we calculate the diffraction pattern using only one of the three magnetic domains, which are shown in the Appendix in Fig.~\ref{Fig:domanSFs} and in Ref.~\cite{Islam2023a}), our model calculations reproduce only two of the six peaks seen experimentally above. The two peaks are distinct for each domain (see Appendix~\ref{Appn:SFDomains}). Therefore, to account for the experimental diffraction pattern, we summed up the contributions from all three domains as shown in Fig.~\ref{Fig:SF_base}\hyperlink{Fig:SF_base}{(d)}. Then we find a remarkable agreement with the experimental observations. The model calculation does not only reproduce the six magnetic peaks around each integer reciprocal-lattice point, but also shows the absence of intensity around the specific three points mentioned above. This firmly establishes the existence of three, symmetry-related magnetic domains predicted previously~\cite{Islam2023a}. 

As noted above, our previous neutron diffraction studies found seven MSGs to be consistent with the data~\cite{Islam2023a}. With the additional data obtained here and using the calculations for the three domains, as illustrated in Fig.~\ref{Fig:SF_base}, we can further narrow down the possibilities. Specifically, our analysis establishes that only the MSGs $P_Ac$, $P_Cm$, and $P_S1$ are consistent with the experimental data, while we can now eliminate the four MSGs $P_A2/c$, $P_C2/m$, $P_C2$, and $P_S\Bar{1}$.

\begin{figure}
\centering
\includegraphics[width=0.9\linewidth]{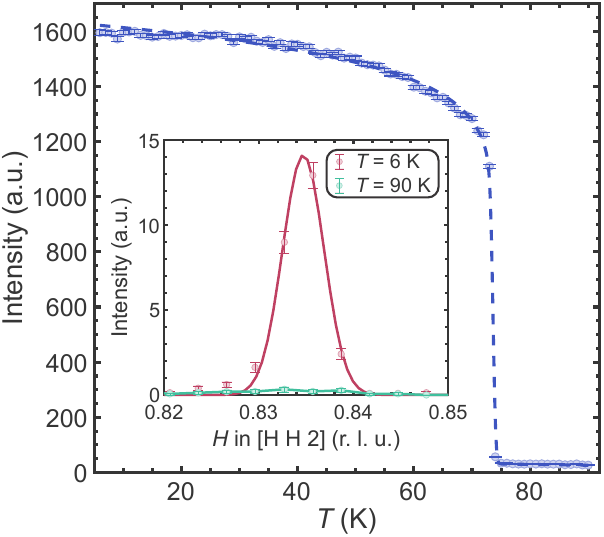}
\caption{The integrated intensity of the prominent magnetic Bragg reflection at $(H,K,L) = (1/6,1/6,2)$~r.l.u.~as a function of temperature $T$, confirming the first-order phase transition at $T_{\rm N}= 70(1)$~K. The dashed line is a fit to the data below $T_{\rm N}$ using a power-law function, $I\sim(1-T/T_{\rm N})^{2\beta}$, that yields $\beta=0.042(6)$. The inset shows the (1/6, 1/6, 2) magnetic Bragg reflection at $T=6$~K and $90$~K.}
\label{Fig:OP}
\end{figure}
To further analyze the temperature behavior of the scattering peaks, Figure~\ref{Fig:OP} shows the integrated intensity of the prominent magnetic Bragg reflection at $(1/6,1/6,1)$ as a function of temperature.  We find a smooth and moderate decrease in intensity from the base temperature to $T_{\rm N}$, at which the intensity abruptly vanishes. This smooth behavior is a notable improvement compared to previously reported results that exhibited slight unexplained anomalies in the temperature dependence below  $T_{\rm{N}}$~\cite{Islam2023a}. The dashed line in Fig.~\ref{Fig:OP} is a fit to the data below $T_{\rm N}$ using a power-law function, $I\sim(1-T/T_{\rm N})^{2\beta}$, yielding an exponent $\beta=0.042(6)$. The small value of $\beta$ has traditionally been associated with a first-order transition, as has been reported for this system from heat-capacity measurements~\cite{Sangeetha2021}. One can understand the first-order nature of the magnetic transition by noting that concurrent with magnetic order the system develops long-range three-state Potts-nematic bond order at $T_{\text{N}}$~\cite{Mulder2010a, Islam2023a}. Since the Potts transition is expected to be first-order in three-dimensions, it can drive the magnetic transition to become first-order, a scenario known to occur in Fe$_{1/3}$NbSe$_2$~\cite{Little2020}. 
We note that an alternative explanation for the experimental observation based on the fact that
magnetic systems with competing interactions and fractal characteristics (due to strong disorder, for example) could exhibit a second-order transition with such a small $\beta$~\cite{Middleton2002}, contrary to the heat-capacity data \cite{Sangeetha2021}. 
While there is currently no direct evidence of fractal behavior in proximity to the transition in \CaMnP\, exploring this possibility by imaging techniques remains intriguing.

\subsection{Experimental Evidence for Spiral Spin Liquid State} 
\begin{figure*}
\centering
\includegraphics[width=\linewidth]{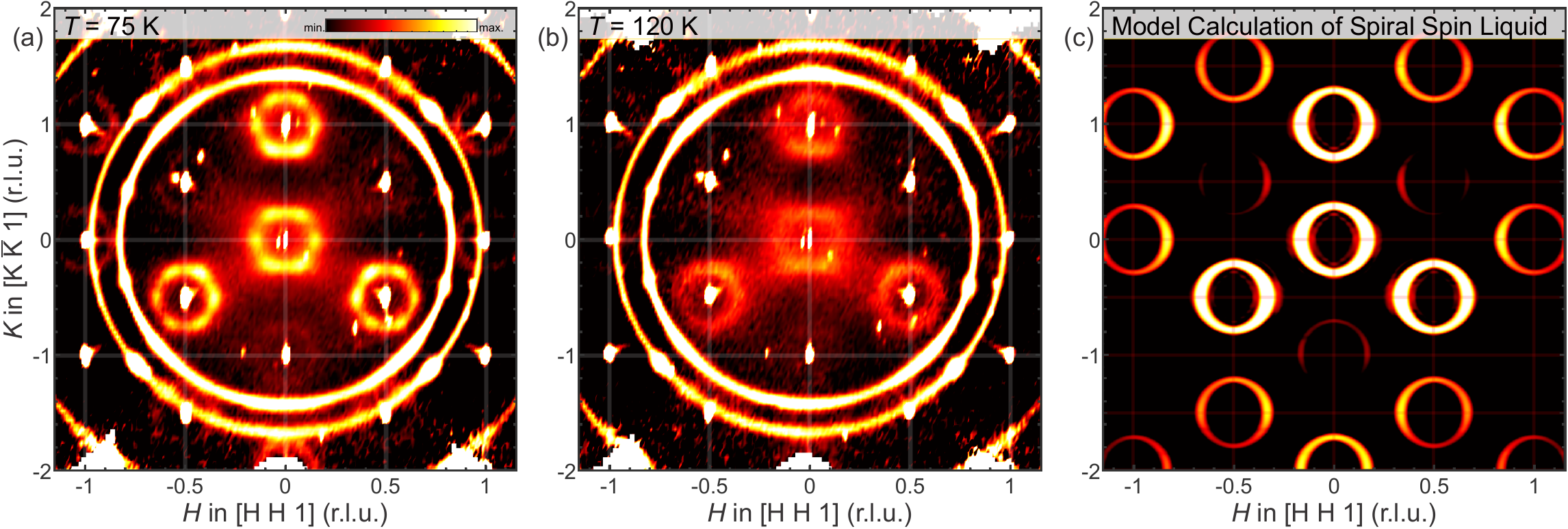}
\caption{(a) and (b) Diffraction patterns in the ($H$,$K$,1) plane above \TN\ at (a) $T=75$ and (b) $120$~K  showing continuous rings of scattering around nuclear Bragg reflections ($\Gamma$ points). While maxima at the Bragg peak locations are still visible at $T=75$~K, the intensity becomes homogeneous along the ring as $T$ increases. The ring has a high degree of circularity (see Fig.~\ref{Fig:SSLcircularity} for a quantitative analysis) and we only observe a small three-fold symmetric deformation of the ring due to the underlying lattice. The larger rings of scattering are due to diffraction from the polycrystalline Al sample holder. 
(c) Model calculations showing the average structure factor of $\approx 1000$ degenerate, single-Q spiral ground state spin configurations. The resulting diffraction pattern is characteristic of the cooperative paramagnetic SSL state.}
\label{Fig:SF_75K}
\end{figure*}

Next, we present diffraction results above the magnetic transition temperature that establish a $U(1)$-symmetric SSL regime with a high degree of circular symmetry in \CaMnP. Figure~\ref{Fig:SF_75K}\hyperlink{Fig:SF_75K}{(a)} shows the diffraction pattern in the $(H,K,1)$ plane at $T = 75$~K, above $T_{\text{N}}$, where the six Bragg reflections observed at low temperatures converge to form continuous rings. The maximum intensity still occurs at the location of the Bragg peaks, but the broadening of the peaks is highly anisotropic with most of the peak weight distributed along the ring. The intensity along the ring becomes more homogeneous as temperature increases, as seen in Fig.~\ref{Fig:SF_75K}\hyperlink{Fig:SF_75K}{(b)} at $T=120$~K. The diffraction maxima form almost perfect circularly symmetric rings with only a small degree of three-fold anisotropy due to the underlying lattice. We quantitatively assess the circularity in Fig.~\ref{Fig:SSLcircularity} of the Appendix.

These rings of high scattering intensity are the primary characteristic of a SSL state and indicate (almost) degenerate magnetic states with wave vectors independent of their $ab$-plane orientation. This demonstrates that the sought-after $XY$-type SSL state with $U(1)$ symmetry in momentum-space is realized in \CaMnP. 

To further confirm the discovery of a SSL in \CaMnP, we first note that the emergence of a $U(1)$-symmetric SSL in the $J_1$-$J_2$-Heisenberg model on the honeycomb lattice for $J_2/J_1 > 1/6$ is well known~\cite{Okumura2010}. Previous work has placed \CaMnP\ into this strongly frustrated region and estimated $J_2/J_1 \approx 0.25$~\cite{Islam2023a}. Thus, to naively model the neutron-diffraction results, we average the calculated diffraction intensity over spin states sampled from the degenerate ground state manifold obtained for $J_2/J_1=0.25$ and $J_3=0$. We thus consider magnetic configurations of the form in Eq.~\eqref{eq:spin_parametrizatio} with wave vectors $\bfqq$ taken from the degenerate, ring-shaped manifold in momentum-space, defined by the solutions of Eq.~\eqref{eq:wavevectors} with $0\leq Q_a,Q_b\leq 2\pi$.
We draw $\approx 1000$ different spin state configurations that are parametrized by wave vectors $\bfqq$ from the degenerate ground state manifold obtained for $J_2/J_1=0.25$ and $J_3=0$. 
In Appendix~\ref{Appn:SSLStates}, we explicitly show a few of these degenerate spin configurations. 
Then we calculate their average structure factor for scattering and show the result in Fig.~\ref{Fig:SF_75K}\hyperlink{Fig:SF_75K}{(c)}. We find that the model calculations adequately capture the essential features of the neutron diffraction data, confirming the SSL state in the highly frustrated regime $J_2/J_1 \approx 0.25$. 

\subsection{Fluctuation-induced SSL} 
\label{subsec:analytical_modeling}
To obtain a quantitative comparison between theory and experiment, we calculate the finite-temperature neutron-scattering cross section $\frac{d \sigma(\bfq,T)}{d \Omega}$ of the frustrated Heisenberg model in Eq.~\eqref{eq:Hspin} (see Appendix~\ref{appendix:subsec:analytical_modeling} for details). 
\begin{figure*}[t!]
\centering
\includegraphics[width=\linewidth]{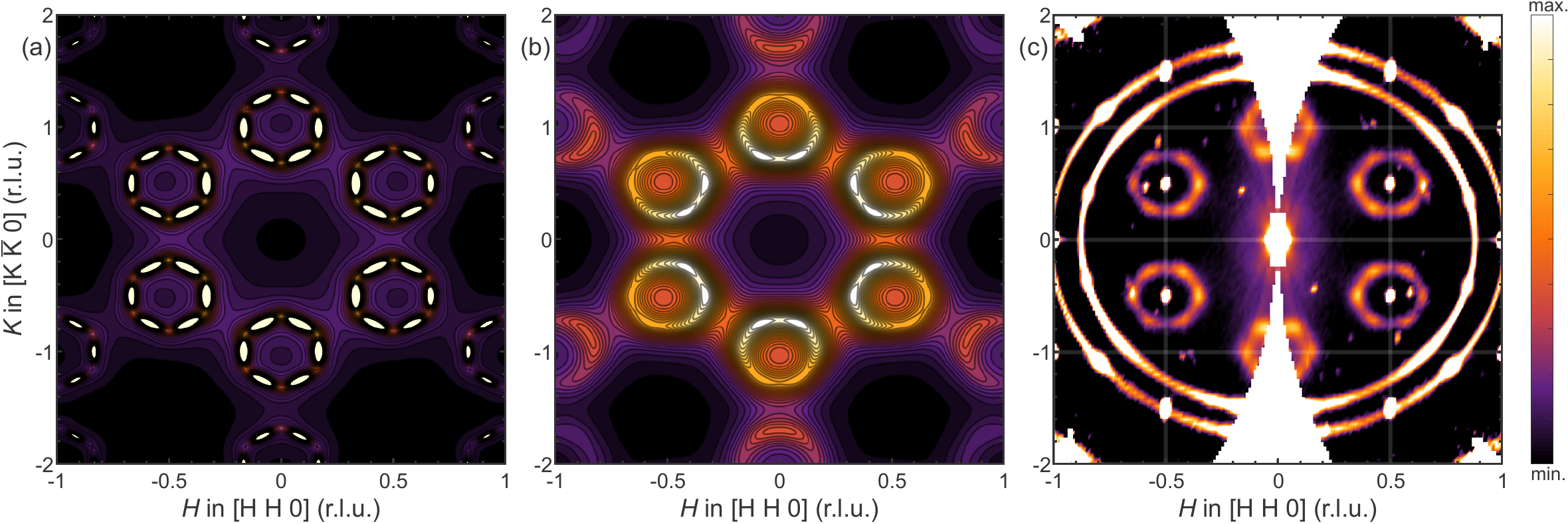}
\caption{(a) Normalized neutron-scattering cross section, $\frac{d \sigma(\bfq, T)}{d \Omega}$ in the ordered phase for $\bfq$ in the $(H,K,0)$ scattering plane. The result is obtained from analytical theory up to quadratic order in spin fluctuations [see Appendix~\ref{appendix:subsec:analytical_modeling}] and for $J_2/J_1 = 0.275$, $J_3/J_1 = 0.05$ and assuming $XY$ spins. The form factor of Mn$^{2+}$ is included and note that the plotted function is independent of $T$ and $J_c$. The main finding is that thermal fluctuations lead to anisotropic broadening of the Bragg peaks. (b) Cross section $\frac{d \sigma(\bfq, T)}{d \Omega}$ in the SSL state from Eq.~\eqref{eq:Iab_high_T} at 
$\Delta(T = 0.2J_1) = 0.14 J_1$. The scattering plane and spin exchange parameters are the same as in panel (a). We clearly find the emergence of the characteristic rings of scattering that are slightly anisotropic due to the underlying threefold-symmetric lattice structure. Comparison with experimental results at $T_{\text{N}} = 75$~K in panel (c) demonstrates excellent agreement and confirms the discovery of a $U(1)$-symmetric SSL state in \CaMnP.}
\label{fig:theory_experiment}
\end{figure*}
First, we consider the impact of thermal spin fluctuations in the ordered phase. As shown in Fig.~\ref{fig:theory_experiment}\hyperlink{fig:theory_experiment}{(a)}, the main effect of the fluctuations is to lead to an anisotropic broadening of the Bragg peaks. Here, we have assumed an equal population of the three symmetry-related domains, set $J_2/J_1 = 0.275$, $J_3/J_1 = 0.05$ and considered a sufficiently large value of $J_c$ and $D_z$ such that different honeycomb planes are ferromagnetically aligned and spins lie in the $XY$ plane. The broadened Bragg peaks are elliptical and extend further in the angular direction. This directly reflects the presence of a massively degenerate spiral ground states for $J_3 = 0$. Even though the degeneracy is weakly broken by finite $J_3$, we find that spins fluctuate more strongly among these spiral states, {\it i.e.}, within the manifold that is degenerate at $J_3 = 0$, than in the perpendicular (radial) direction. Experimentally observing such anisotropic fluctuations by performing a detailed scattering study just below the ordering temperature $T \lesssim T_\text{N}$ is an interesting direction to explore in the future. 
To study the spin correlations in the paramagnetic regime ($T>T_{\text{N}}$), we use the self-consistent Gaussian approximation (SCGA), which yields~\cite{Bergman2007}
\begin{equation}
     I_{\alpha\beta}(\mathbf{k}) \equiv \left\langle\mathbf{S}_{\mathbf{k},\alpha}\cdot\mathbf{S}_{-\mathbf{k},\beta}\right\rangle=2T\left[\hat{J}(\mathbf{k})+\Delta(T)\sigma_0\right]_{\alpha,\beta}^{-1} \,.
    \label{eq:SkSk}
\end{equation}
Here, $\hat{J}(\mathbf{k})$ is the Fourier transform of the exchange matrix of the spin model in Eq.~\eqref{eq:Hspin}. The gap function $\Delta(T)$ is determined self-consistently using the spin length constraint as
\begin{equation}
    \frac{1}{T}=\sum\limits_{j=0,1}\int\frac{d^3q}{\left(2\pi\right)^3}\frac{1}{\kappa_j(\mathbf{q})+\Delta(T)} \,.
    \label{eq:Delta}
\end{equation}
Here, $\kappa_{0,1}(\mathbf{q})=\epsilon_{\mp}(\mathbf{q})-E_0$ with $\epsilon_{\mp}(\mathbf{q})$ denoting the eigenvalues of the exchange matrix and $E_0$ representing the classical ground state energy. Given a set of exchange parameters $J_1$, $J_2$, $J_3$ and $J_c$, we solve Eq.~\eqref{eq:Delta} numerically and substitute the resulting $\Delta(T)$ into Eq.\eqref{eq:SkSk}, which is more readily written in the eigenbasis of $\hat{J}(\mathbf{q})$ as 
\begin{equation}
    I_{\alpha\beta}(\mathbf{q},T)=2T\left[\frac{\hat{U}_{\alpha1}^{\dag}(\mathbf{q})\hat{U}_{1\beta}(\mathbf{q})^{\null}}{\kappa_0(\mathbf{q})+\Delta(T)}+\frac{\hat{U}_{\alpha2}^{\dag}(\mathbf{q})\hat{U}_{2\beta}(\mathbf{q})^{\null}}{\kappa_1(\mathbf{q})+\Delta(T)}\right]\,,
    \label{eq:Iab_high_T}
\end{equation}
where $\hat{U}(\mathbf{q})$ diagonalizes $\hat{J}(\mathbf{q})$.

In Fig.~\ref{fig:theory_experiment}\hyperlink{fig:theory_experiment}{(b)}, we show the resulting scattering cross-section $\frac{d \sigma(\bfq, T)}{d \Omega}$ in the $(H,K,0)$ plane, obtained with Eq.~\eqref{eq:Iab_high_T} at temperature $T = 0.2 J_1/k_{\rm B}$, which corresponds to $\Delta(T=0.2J_1/k_{\rm B})=0.14J_1/k_{\rm B}$. The theoretical result shows excellent agreement with the experimental diffraction result at $T = 75$~K, which is seen in Fig.~\ref{fig:theory_experiment}\hyperlink{fig:theory_experiment}{(c)}. Both panels exhibit six characteristic rings of scattering that are centered around nuclear peaks (only seen in the experimental data). Since $T = 75$~K is just above the phase transition into the paramagnetic SSL state, the intensity around the rings still shows maxima at the location of the magnetic Bragg peaks, which is also reflected in the theoretical result. The intensity across the ring becomes more homogeneous with increasing temperature, which is shown in the experimental data at $T=120$~K in Fig.~\ref{Fig:SSLcircularity} and for the theory in Fig.~\ref{fig:temperature_dependence_I}.

\subsection{Classical Spin Dynamics Simulations}
\label{subsec:simulations}
\begin{figure}
\centering
\includegraphics[width=\linewidth]{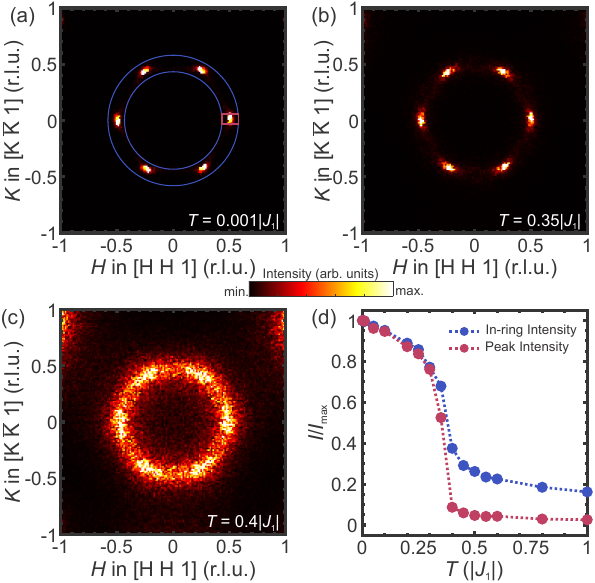}
\caption{Structure factor in ($H$,$K$,1) reciprocal plane from spin dynamics simulations at finite temperature for anisotropic $J_1$-$J_2$-$J_3$-$D_z$ Heisenberg model. Model parameters are set to $J_1 = 1$, $J_2 = 0.275$, $J_3 = 0.05$ and $D_z = 0.05$ such that the ground state is a spin spiral with wavevector $H = K = \frac16$ (or symmetry related wavevectors). Panels (a-c) show results for three temperatures $T/J_1 = \{10^{-3}, 0.35, 0.4\}$ as indicated. 
The colormaps are symmetrized with respect to 3-fold symmetry to improve the statistics in the obtained Fourier maps. Panel (d) shows the temperature dependence of the normalized integrated intensity. Blue curve: integration inside the constant-wavelength ring; red curve: integration in close vicinity of the $\mathbf{Q}=(1/6, 1/6, 1)$ peak. Both integration areas are included in panel (a).}
\label{Fig:sunny1}
\end{figure}
To further validate the applicability of the $J_1$-$J_2$-$J_3-D_z$ Heisenberg model in reproducing the emergence of the magnetic ground and the SSL states in \CaMnP\, we conduct classical spin dynamics numerical simulations using the \textsc{Su(n)ny} program package~\cite{Zhang2021}. These simulations treat the Mn spins as classical dipoles, accounting for magnetic couplings up to the third-nearest neighbor. Exchange couplings, anisotropy, and temperature are normalized to the antiferromagnetic nearest-neighbor interaction $J_1$; $J_2=0.25(J_1+2J_3); J_3=0.05J_1$. These exchange parameters agree with our previous analysis above and in Ref.~\cite{Islam2023a}. 
In accordance with experimental results, an easy-plane anisotropy $D_z=0.05J_1$ is introduced to force the spins to lie in the $ab$-plane. Based on crystal-structure and experimental observations, \CaMnP\ exhibits quasi-two-dimensional magnetism. This allows us to simulate a single bilayer system created by two triangular sublattices with a large $ab$-plane supercell size of 100$\times$100 crystallographic unit cells. The structure factor at finite temperatures is obtained (by following similar methodology outlined in Ref.~\cite{Islam2023a} for each spin configuration) through a preliminary thermalization process driving the system to thermal equilibrium. During the thermalization, spin configurations are sampled using Langevin dynamics~\cite{Dahlbom2022} with a minimum of $2 \times 10^4$ time steps per spin for each temperature. The temperature of the Langevin sampler is gradually decreased to simulate the annealing process during thermalization. This comprehensive approach is in agreement with our analytical theory and allows further assessing the model's efficacy in capturing the observed SSL state in \CaMnP\ under realistic conditions. The parameters used to predict the cycloidal and degenerate ring magnetic configurations effectively capture the key features observed in our experimental results.

\section{Conclusions and Outlook}
\label{Sec:Conclu} 
Neutron-diffraction studies of \CaMnP\ confirm the previously observed magnetic cycloidal ground state with $6 \times 6$ magnetic unit cell, signaling extended spin exchange couplings with substantial frustration in the corrugated $S=5/2$ honeycomb compound. Agreement between simulations using a frustrated $J_1$-$J_2$-$J_3$-$D_z$ Heisenberg honeycomb model with $J_2 \approx J_1/4 \gg J_3$ and the observed neutron-scattering in a large reciprocal-space volume demonstrate consistency with the complex magnetic order described by one of the MSGs $P_Ac$, $P_Cm$, or $P_S1$. We directly observe the presence of three equivalent magnetic domains, each rotated by 120 degrees with respect to one another. The observation of threefold rotational symmetry breaking provides further support to explain the unusual, strong first-order magnetic transition in \CaMnP~\cite{Sangeetha2021} in terms of a three-state Potts-nematic transition in the system~\cite{Islam2023a}. 

Above the N\'eel temperature $T_{\rm N}$, we find strong experimental evidence for the emergence of a $U(1)$-symmetric SSL state by observing the formation of rings of scattering in reciprocal-space. This development of such a highly degenerate, cooperative paramagnetic spin liquid state marks a significant transition in magnetic behavior of the system. 
Remarkably, \CaMnP\ thus seems to realize the sought-after $U(1)$-symmetric SSL phase of planar $XY$ spins for which topologically protected momentum-space vortices are predicted to play a key role~\cite{Yan2022a, Gonzalez2024}. These exotic excitations describe windings of the coarse grained ordering wavevector $\bfqq(\bfr)$ around the circularly shaped degenerate ground state manifold and underlie the description of the low-energy fluctuations in the SSL phase in terms of a tensor gauge theory of fractons. 
Using \CaMnP, further experimental research into such exotic theories will be achievable. One intriguing open question is whether the transition into the SSL phase corresponds to a Kosterlitz-Thouless (KT) unbinding transition of topological momentum vortices~\cite{Gonzalez2024}. 

Other interesting future research avenues are the exploration of the dynamic nature of the ground state, its short-range spiral spin configurations and their role in collective magnetic behaviors. Specifically, inelastic neutron-scattering could be used to precisely determine the spin exchange coupling and anisotropy parameters and to search for signatures of SSL and neighboring phases in the dynamic response. Theoretical studies of a square-lattice SSL predict intense streaks of scattering should emerge from the spiral wavevectors in the wavevector-frequency plane~\cite{Gonzalez2024} that can distinguish the SSL from nearby phases. To directly confirm the presence of momentum-space vortices requires extracting the local spin structure using, for example, Lorentz transmission electron microscopy (TEM). This would give access to the coarse grained ordering momentum field $\bfqq(\bfr)$, whose derivatives define an emergent tensorial electric field. The electric field correlator  is predicted to exhibit unique fourfold pinch points due to a generalized Gauss' law that is characteristic of the SSL phase~\cite{Yan2022a, Gonzalez2024}. 

Another fundamental open question to address using \CaMnP\ is how the SSL transforms at higher temperature and whether it disappears via a phase transition or a crossover. If the radius of the ring is sufficiently small, the SSL is predicted to transform into a ``pancake spin liquid" as $T$ increases~\cite{Okumura2010, Yan2022a}. The pancake liquid exhibits a structure factor with a high-intensity plateau for wavevectors inside the ring. This state has not yet been observed experimentally. Even though the ring radius is not particularly small in \CaMnP, the fate of the SSL state in \CaMnP\ with increasing temperatures presents an intriguing direction to explore. Finally, probing \CaMnP\ in an applied magnetic field could explore a potentially rich phase diagram. For perpendicular field directions, theoretical studies predict the emergence of different single-Q and double-Q phases, while the behavior under in-plane fields is largely unexplored. Given that the characteristic field scale is $H_c \approx J_1/g \mu_\text{B} S \approx 16$~T, exploring the rich phase diagram in a magnetic field is feasible under normal laboratory conditions and offers a promising path for future discovery. 

\acknowledgments
P.P.O. acknowledges valuable discussions with J.~Reuther. 
This research was supported by the U.S. Department of Energy, Office of Basic Energy Sciences, Division of Materials Sciences and Engineering. Iowa State University operates Ames National Laboratory for the U.S. Department of Energy under Contract No.~DE-AC02-07CH11358. This work was performed in part at Aspen Center for Physics, which is supported by National Science Foundation grant PHY-2210452.


\appendix

\section{Structure Factors from Each Domain}
\label{Appn:SFDomains}
\begin{figure*}
\includegraphics[width=0.8\linewidth]{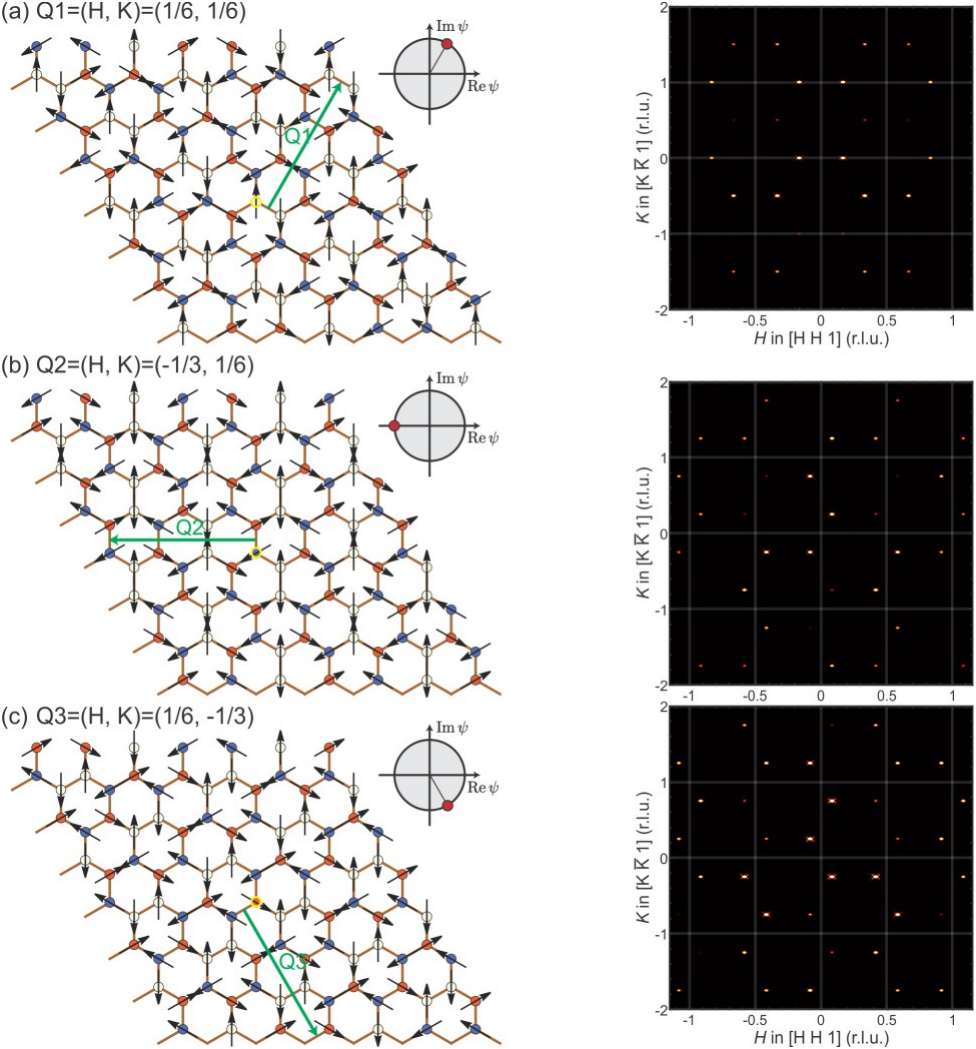}
\caption{(a) to (c): The three cycloidal magnetic domains and their corresponding structure factor calculations. These structure factor calculations are summed up to model the diffraction pattern in Fig.~\ref{Fig:SF_base}(c) producing Fig.~\ref{Fig:SF_base}(d).}
\label{Fig:domanSFs}
\end{figure*}

In Fig.~\ref{Fig:domanSFs}, we show diffraction calculation results for each of the three possible magnetic domains described by the wavevectors $(H,K,L) = (\frac16, \frac16, m)$ (integer $m$) and symmetry related. The experimentally observed diffraction pattern consists of an equal superposition of the three diffraction patterns obtained for the individual domains. The superposition is shown in Fig.~\ref{Fig:SF_base}\hyperlink{Fig:SF_base}{(d)}.

\section{Examples of Degenerate Configurations in the SSL}
\begin{figure*}
\includegraphics[width=0.65\linewidth]{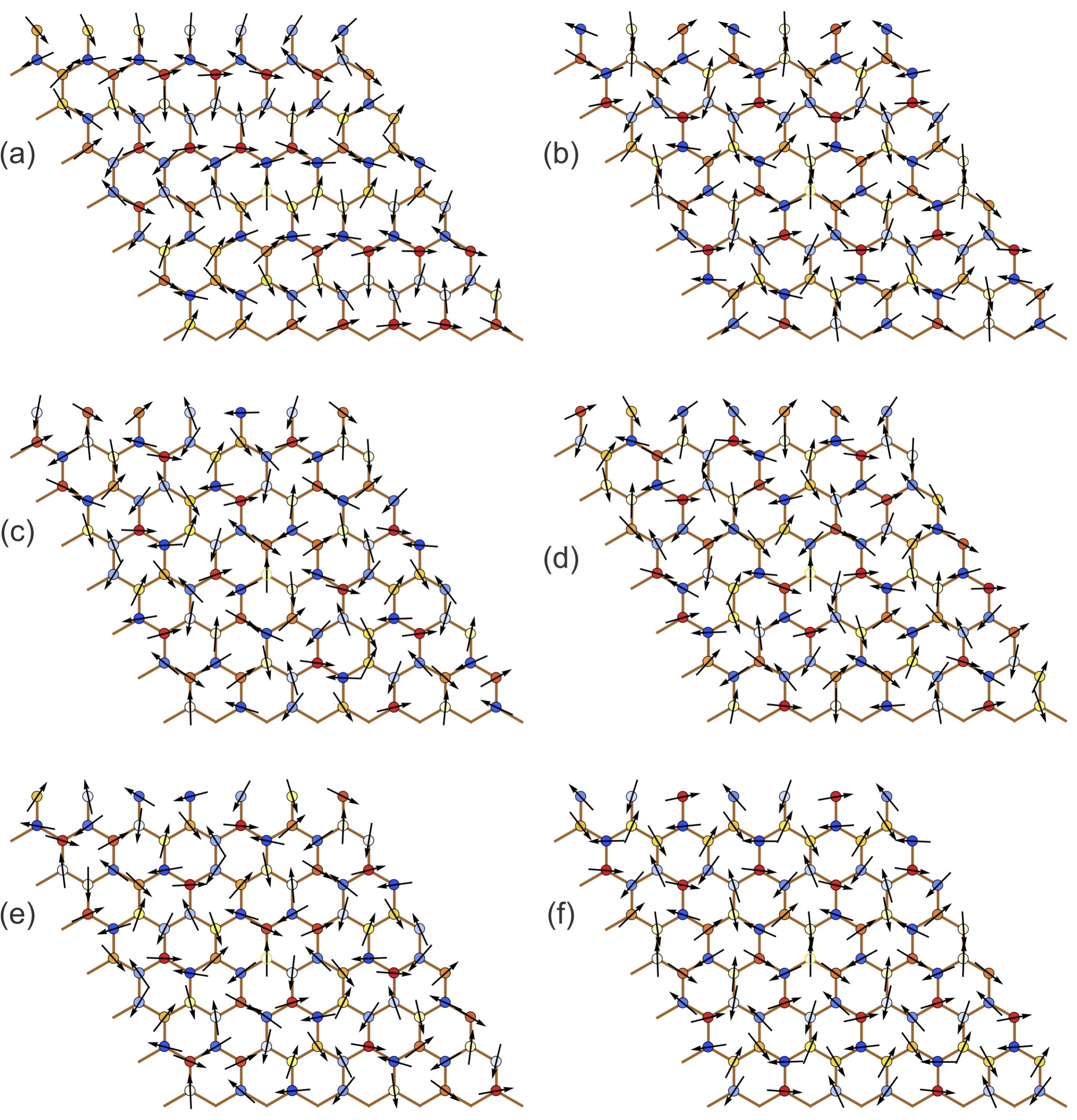}
\caption{Six examples of magnetic configurations that belong to the continuum of degenerate states at $J_2 = J_1/4$ and $J_3 = 0$ that give rise to the rings of scattering above $T_{\rm N}$. 
}
\label{Fig:SSLexamples}
\end{figure*}

In Fig.~\ref{Fig:SSLexamples}, we explicitly show six different single-Q spin configurations in the ground state with $J_2/J_1 = 0.25$ and $J_3 = 0$.  
\label{Appn:SSLStates}

\section{Circularity of Magnetic Spiral Spin Liquid Scattering}
\begin{figure}
\includegraphics[width=\linewidth]{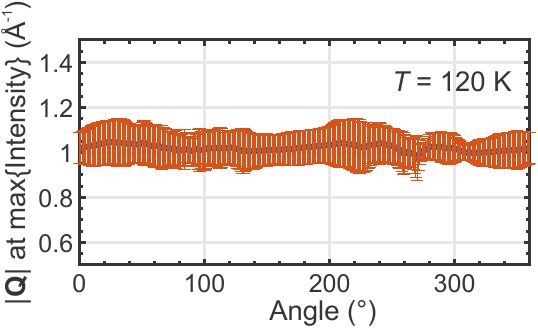}
\caption{The plot shows the variation of the magnitude of the scattering vector $|\bf{Q}|$ at the maximum intensity as a function of the azimuthal angle for circular magnetic scattering of the SSL phase in Fig.~\ref{Fig:SF_75K}(b) at $T=120$ K. The nearly-constant $|\bf{Q}|$ value with minor angular fluctuations indicates a high degree of circular $U(1)$ symmetry in the SSL state.}
\label{Fig:SSLcircularity}
\end{figure}

We assess the circularity of the magnetic SSL scattering by performing line-cuts as a function of $Q$ across all angles in Fig.~\ref{Fig:SF_75K}(b). To minimize interference from the high-intensity region at the center and scattering from other rings, we limit the line-cuts to the range of $Q = 0.95$ to $1.3~{\rm \AA}^{-1}$. The peak positions are then determined by fitting a Gaussian distribution to each cut, identifying the $Q$ value corresponding to the maximum intensity. Figure~\ref{Fig:SSLcircularity} presents the extracted $Q$ values, with the Gaussian fit widths reported as error bars.

\section{Analytical Modeling} 
\label{appendix:subsec:analytical_modeling}
To make a more quantitative comparison between theory and experiment, we now perform an analytical description of the SSL in the $J_1$-$J_2$-$J_3$-$J_c$-$D_z$ Heisenberg model of Eq.~\eqref{eq:Hspin}. To directly compare theory to experiment, we calculate the neutron-scattering cross section
\begin{equation}
    \frac{d \sigma(\bfq,T)}{d \Omega} = C f^2(q) \sum_{\alpha, \beta \atop a,b} \left( \delta_{ab} - \frac{q_a q_b}{|\bfq|^2}\right) \left\langle S_{\alpha,a}(\bfq) S_{\beta, b}(-\bfq) \right\rangle \,.
\end{equation}
Here, $C$ is a constant and $f(q)$ is the form factor of Mn$^{2+}$ ions, 
\begin{equation}
    f(q)=\sum\limits_{j=1}^{4}A_je^{-p_j\left(\frac{q}{4\pi}\right)^2}\text{ ,}
\end{equation}

\noindent with $A_1=0.4220$, $A_2=0.5948$, $A_3=0.0043$, $A_4=-0.0219$, $p_1=17.6840$\,\AA$^2$, $p_2=6.0050$\,\AA$^2$, and $p_3=-0.6090$\,\AA$^2$. Besides, $a,b = x,y,z$ denote spin components and $\alpha, \beta = A,B$ sublattice indices. Note that for our model, we find that only $a=b$ correlations are nonzero and $\left\langle S_{\alpha,x}(\bfq) S_{\beta, x}(-\bfq) \right\rangle = \left\langle S_{\alpha,y}(\bfq) S_{\beta, y}(-\bfq) \right\rangle$. This simplifies the expression for the scattering cross section to
\begin{equation}
    \frac{d \sigma(\bfq,T)}{d \Omega} = C f^2(|\bfq|) \left( 1 + \frac{q_z^2}{|\bfq|^2}\right) I(\bfq,T) \,,
\end{equation}
where we have introduced the total spin structure factor 
\begin{equation}
    I(\mathbf{q},T) = I_{AA}(\mathbf{q},T) + I_{BB}(\mathbf{q},T) + 2 \text{Re} \, I_{AB} (\mathbf{q},T)
\end{equation}
that is given in terms of the sublattice-resolved spin structure factor 
\begin{align}
    I_{\alpha\beta}(\mathbf{q},T)&=\int d\mathbf{r}_i\int d\mathbf{r}_j\,e^{-i\mathbf{q}\cdot\left(\mathbf{r}_i -\mathbf{r}_j\right)}\Bigl[\bigl\langle\mathbf{S}_{i,\alpha}\cdot \mathbf{S}_{j,\beta}\bigr\rangle \nonumber \\ &- \bigl\langle\mathbf{S}_{i,\alpha}\bigr\rangle\cdot\left\langle\mathbf{S}_{j,\beta}\right\rangle\Bigr]\text{.}
    \label{eq:Iab}
\end{align}
Here, $\left\langle\cdots\right\rangle=\text{tr}\left(e^{-\beta H}\cdots\right)/\mathcal{Z}$ denotes the thermal average with inverse temperature $\beta = 1/T$ and $\mathcal{Z}=\text{tr}\left(e^{-\beta H}\right)$ is the partition function. Besides, $\mathbf{r}_i=\mathbf{R}_i+\boldsymbol{\tau}_{\alpha}$ denotes the physical position of the site belonging to the $\alpha$-sublattice in the unit cell $i$ located at $\mathbf{R}_i$. We choose the basis vectors to be $\boldsymbol{\tau}_{A}=\mathbf{0}$ and $\boldsymbol{\tau}_{B}=a\left(0,\frac{1}{\sqrt{3}},z_{AB}\right)$, with $z_{AB}=0.25a_c$ being the spacing between the two Mn layers in the unit cell. In the following, we first analyze the effect of thermal spin fluctuations in the ordered phase below \TN ~, before considering the paramagnetic SSL regime above \TN. 

\subsubsection{Anisotropic Thermal Fluctuations at Low Temperatures}
Here we consider the impact of thermal spin fluctuations on the ordered phase in the low-temperature regime $T<T_{\text{N}}$. 
The main effect of the fluctuations is to vary the spin orientation away from the position in the ground state, without modifying their amplitude. To model thermal fluctuations we can thus start from a single-Q ground state spin configuration described by Eq.~\eqref{eq:spin_parametrizatio} and replace the ground state angles $\theta_{i,\alpha}^{(0)}=\mathbf{Q}\cdot \mathbf{R}_i+(\phi + \pi )\delta_{\alpha,B}$ by
\begin{equation}
\theta_{i,\alpha}=\theta_{i,\alpha}^{(0)}+\psi_{i,\alpha}\text{ .}
\label{eq:SMtheta}
\end{equation}
The variables $\psi_{i,\alpha}$ thus characterize the deviations from a ground state configuration. Assuming that the $\psi_{i,\alpha}$ are randomly distributed according to a Gaussian distribution, we expand the structure factor in Eq.~\eqref{eq:Iab} up to second order in $\psi_{i,\alpha}$ and evaluate the resulting Gaussian integrals over $\psi_{i, \alpha}$, to obtain 
\begin{align}
    I_{\alpha\beta}(\mathbf{q})=TS^2&\left[e^{-i\Lambda_{\alpha\beta}(\mathbf{Q})}\hat{\chi}_{\alpha\beta}(\mathbf{Q}-\mathbf{q})\right.+\nonumber\\
    &\left. e^{i\Lambda_{\alpha\beta}(\mathbf{Q})}\hat{\chi}_{\alpha\beta}(-\mathbf{Q}-\mathbf{q})\right]
    \label{eq:Iq_low_T}
\end{align}
where
\begin{equation}
    \Lambda_{\alpha\beta}(\mathbf{Q})=\mathbf{Q}\cdot\left(\boldsymbol{\tau}_{\alpha}-\boldsymbol{\tau}_{\beta}\right)+\left(\pi+\phi\right)\left(\delta_{\alpha,A}\delta_{\beta,B}-\delta_{\alpha,B}\delta_{\beta,A}\right) \text{ .}
\end{equation}

\noindent Furthermore, $\hat{\chi}_{\alpha\beta}(\mathbf{q})$ are the matrix elements of the spin susceptibility
\begin{equation}
    \hat{\chi}(\mathbf{q})=\begin{pmatrix}
        f(\mathbf{q}) & g(\mathbf{q})\\
        g(-\mathbf{q}) & f(\mathbf{q})
    \end{pmatrix}^{-1}
    \label{eq:SMchi}
\end{equation}

\noindent with
\begin{align}
    &f(\mathbf{q})=J_1\sum\limits_{\boldsymbol{\delta}_1}\cos(\mathbf{Q}\cdot\boldsymbol{\delta}_1+\phi+\pi)\nonumber\\
    &+J_2\sum\limits_{\boldsymbol{\delta}_2}\cos(\mathbf{Q}\cdot\boldsymbol{\delta}_2)\left(1-e^{-i\mathbf{q}\cdot\boldsymbol{\delta}_2}\right)\nonumber\\
    &+J_3\sum\limits_{\boldsymbol{\delta}_3}\cos(\mathbf{Q}\cdot\boldsymbol{\delta}_3+\phi+\pi)\text{ ,}
\end{align}
\noindent and 
\begin{align}
    &g(\mathbf{q})=-J_1\sum\limits_{\boldsymbol{\delta}_1}\cos(\mathbf{Q}\cdot\boldsymbol{\delta}_1+\phi+\pi)e^{-i\mathbf{q}\cdot\left(\boldsymbol{\delta}_1+\boldsymbol{\tau}_B-\boldsymbol{\tau}_A\right)}\nonumber\\
    &-J_3\sum\limits_{\boldsymbol{\delta}_3}\cos(\mathbf{Q}\cdot\boldsymbol{\delta}_3+\phi+\pi)e^{-i\mathbf{q}\cdot\left(\boldsymbol{\delta}_3+\boldsymbol{\tau}_B-\boldsymbol{\tau}_A\right)}\text{ .}
\end{align}

\noindent In the previous equations, $\boldsymbol{\delta}_{j}$, $j=1,2,3$ are the vectors connecting first, second and third neighbors in the honeycomb lattice, respectively. Eq.(\ref{eq:SMchi}) is obtained by parametrizing the spins by $\mathbf{S}_{i,\alpha}=\left(\sin\theta_{i,\alpha},\cos\theta_{i,\alpha}\right)$, with $\theta_{i,\alpha}$ given by Eq.(\ref{eq:SMtheta}), substituting it in the spin Hamiltonian defined in Eq.(\ref{eq:Hspin}) and expanding it up to second order in the fluctuations $\psi_{i,\alpha}$.

Note that in the low-temperature regime, the temperature dependence in $I_{\alpha\beta}(\mathbf{q})$ appears as a global factor. Therefore, when normalizing the scattering cross section by its largest value as shown in Fig.\ref{fig:theory_experiment}, $T$ drops out.  

Our diffraction results at $T < T_{\text{N}}$ in Sec.~\ref{subsec:magnetic_structure} indicate that the system exhibits three distinct single-Q domains with symmetry-related wavevectors $(H,K,L)=\left(\frac{1}{6},\frac{1}{6},m\right)$, $\left(-\frac{1}{3},\frac{1}{6},m\right)$ and $\left(\frac{1}{6},-\frac{1}{3},m\right)$ with integer $m$. These wavevectors emerge in the model for $J_2 = \frac14(J_1 + 2 J_3)$ at small antiferromagnetic $J_3>0$~\cite{Islam2023a}. Thus, in
Figure~\ref{fig:theory_experiment}~\hyperlink{fig:theory_experiment}{(a)} we show results for the total structure factor $I(\mathbf{q},T)$ in the $L = 0$ plane, obtained for an equal population of the three symmetry-related domains. We set $J_2/J_1 = 0.275$, $J_3/J_1 = 0.05$, and a sufficiently large value of $J_c$ and $D_z$ such that different honeycomb planes are ferromagnetically aligned and the spins lie in the $XY$ plane. We also include the form factor of Mn$^{2+}$ in our calculations. In contrast to the diffraction in the $L=1$ plane, here the central six peaks around the $H=K=0$ point are absent. This agrees with the experimental results [see also Fig.~\ref{fig:theory_experiment}~\hyperlink{fig:theory_experiment}{(c)}]. 

Importantly, we observe in our theoretical result in Fig.~\ref{fig:theory_experiment}\hyperlink{fig:theory_experiment}{(a)} that thermal spin fluctuations broaden the magnetic Bragg peaks in an anisotropic way. The broadened peaks are elliptical and extend further in the angular direction. This directly reflects the presence of a massively degenerate spiral ground states for $J_3 = 0$. Even though the degeneracy is weakly broken by finite $J_3$, we find that the spins fluctuate more strongly among these spiral states, {\it i.e.}, within the manifold that is degenerate at $J_3 = 0$, than in the perpendicular (radial) direction. Experimentally observing such anisotropic fluctuations by performing a detailed scattering study just below the ordering temperature $T \lesssim T_\text{N}$ is an interesting direction to explore in the future. 

\subsubsection{Spiral Spin Liquid}
Next, we study the behavior of the spin correlations in the high-temperature, paramagnetic regime ($T>T_{\text{N}}$). This analysis yields information about the spin correlations in the SSL phase. Note that $\langle\mathbf{S}_{i,\alpha}\rangle=0$ in the paramagnetic state and Eq.~\eqref{eq:Iab} simplifies to the Fourier transform of the spin correlation function $\langle\mathbf{S}_{i,\alpha}\cdot \mathbf{S}_{j,\beta}\rangle$. To calculate the spin structure factor, we use the self-consistent Gaussian approximation (SCGA) method~\cite{Bergman2007,Conlon2010}. There, one first employs the spherical approximation that relaxes the local constraint on the normalization of individual spins, by introducing a global minimization condition, namely $\sum\limits_{i,\alpha}\left|\mathbf{S}_{i,\alpha}\right|^2=N$, where $N$ denotes the total number of lattice sites. This soft-spin condition reflects the fact that strong spin fluctuations near \TN ~can lead to a suppression of the spin magnitude in addition to the fluctuations in their orientation, and yields~\cite{Bergman2007}
\begin{equation}
    \left\langle \mathbf{S}_{\mathbf{k},\alpha}\cdot\mathbf{S}_{-\mathbf{k},\beta}\right\rangle=2T\left[\hat{J}(\mathbf{k})+\Delta(T)\sigma_0\right]_{\alpha,\beta}^{-1} \,.
    \label{appendix_eq:SkSk}
\end{equation}
Here, $\hat{J}(\mathbf{k})$ is the Fourier transform of the exchange matrix of our spin model in Eq.~\eqref{eq:Hspin}. Note that this is a two-by-two matrix since we have two sublattice degrees of freedom, and $\sigma_0$ is the two-by-two identity matrix. The global factor of two in the previous equation reflects that we consider planar $XY$ spins here with two spin components. The gap function $\Delta(T)$ is determined self-consistently using the spin normalization condition as
\begin{equation}
    \frac{1}{T}=\sum\limits_{j=0,1}\int\frac{d^3q}{\left(2\pi\right)^3}\frac{1}{\kappa_j(\mathbf{q})+\Delta(T)} \,.
    \label{appendix_eq:Delta}
\end{equation}
Here, the function $\kappa_{0,1}(\mathbf{q})=\epsilon_{\mp}(\mathbf{q})-E_0$ with  $\epsilon_{\mp}(\mathbf{q})$ denoting the eigenvalues of the exchange matrix and $E_0$ representing the classical ground state of the single-$Q$ helical ground state. 

For a chosen set of exchange parameters $J_1$, $J_2$, $J_3$ and $J_c$, we solve Eq.~\eqref{eq:Delta} numerically and substitute the resulting $\Delta(T)$ into Eq.\eqref{eq:SkSk}. Note that the spin structure factor in Eq.~\eqref{eq:SkSk} is more readily written in the basis that diagonalizes the exchange matrix, 
\begin{equation}
    I_{\alpha\beta}(\mathbf{q},T)=2T\left[\frac{\hat{U}_{\alpha1}^{\dag}(\mathbf{q})\hat{U}_{1\beta}(\mathbf{q})^{\null}}{\kappa_0(\mathbf{q})+\Delta(T)}+\frac{\hat{U}_{\alpha2}^{\dag}(\mathbf{q})\hat{U}_{2\beta}(\mathbf{q})^{\null}}{\kappa_1(\mathbf{q})+\Delta(T)}\right] \text{ .}
    \label{appendix_eq:Iab_high_T}
\end{equation}
Here, $\hat{U}(\mathbf{q})$ is the matrix that diagonalizes $\hat{J}(\mathbf{q})$.

In Fig.~\ref{fig:theory_experiment}\hyperlink{fig:theory_experiment}{(b)}, we show the total structure factor in the $(H,K,0)$ plane, obtained with Eq.~\eqref{eq:Iab_high_T} at temperature $T = 0.2J_1$. The theoretical result shows excellent agreement with the experimental diffraction result in the same scattering plane at $T = 75$~K, which is included in Fig.~\ref{fig:theory_experiment}\hyperlink{fig:theory_experiment}{(c)}. Both panels exhibit six characteristic rings of scattering that occur around the nuclear peaks (only seen in the experimental data). Since $T = 75$~K is just above the phase transition, the intensity around the rings still shows maxima at the location of the magnetic Bragg peaks, which is also seen in the theoretical result. The intensity across the ring becomes more homogeneous with increasing temperature, which is shown at $T=120$~K in Fig.~\ref{Fig:SSLcircularity} above.

\section{Temperature dependence of structure factor}

\begin{figure*}
    \centering
    \includegraphics[width=\linewidth]{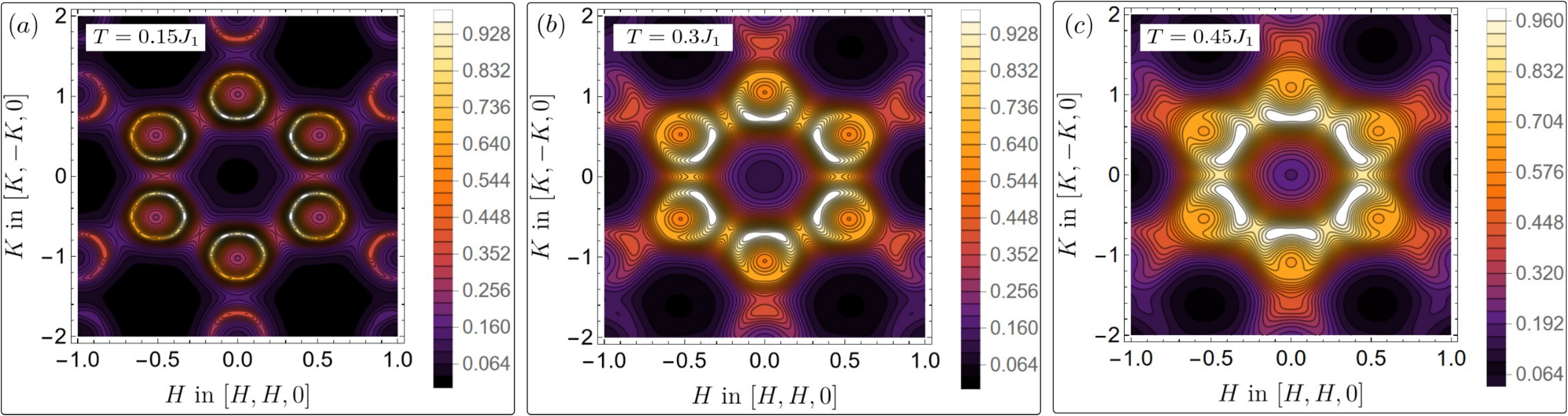}
    \caption{Temperature dependence of the normalized total structure factor $I(\bfq,T)$ in the SSL phase. The different panels are for the indicated temperatures: $T/J_1 = 0.15$ (a), $0.3$ (b), and $0.45$ (c). The results are is obtained from Eq.~\eqref{eq:Iab_high_T} with spin exchange parameters set to $J_2/J_1 0.275$, $J_3/J_1 = 0.05$, and include the form factor for Mn$^{2+}$. The effect of increasing temperature is mostly to broaden the characteristic rings of high scattering intensity in the SSL phase. The structure factor is normalized to its maximum value in each panel.}
    \label{fig:temperature_dependence_I}
\end{figure*}

In Fig.~\ref{fig:temperature_dependence_I}, we show the analytically obtained temperature dependence of the total structure factor $I(\bfq, T)$ in the SSL phase. The figure contains results for $T/J_1= 0.15$, $0.3$, and $0.45$. While increasing temperature leads to a broadening of the characteristic rings of high scattering intensity, the contours can be seen even at the highest temperature, which is more than twice as large as the temperature chosen in Fig.~\ref{fig:theory_experiment}~\hyperlink{fig:theory_experiment}{(b)} of the main text. This demonstrates that, within our approximate analytical theory, the SSL state is very robust to thermal fluctuations, which agrees well with experimental observations. The intensity inside the rings increases with increasing temperature, which is somewhat reminiscent of the pancake liquid phase~\cite{Okumura2010}, even though the maximum intensity still occurs at the circumference of the ring. Our theory does not capture a possible KT transition due to the unbinding of vortices in momentum-space, which was numerically observed (for a different model) in Ref.~\cite{Gonzalez2024}.

\pagebreak
\newpage

\bibliography{main.bbl}

\end{document}